\documentclass[prl,twocolumn,showpacs]{revtex4}
\usepackage{amsmath}
\usepackage[dvips,xdvi]{graphicx}
\usepackage{amssymb}
\usepackage{epsfig}

\catcode`\Š = \active \catcode`\š = \active \catcode`\Ÿ = \active
\catcode`\€ = \active \catcode`\… = \active \catcode`\† = \active
\catcode`\§ = \active \catcode`\Ž = \active \catcode`\ = \active
\catcode`\' = \active \catcode`\™ = \active \catcode`\ = \active
\catcode`\¿ = \active \catcode`\˜ = \active \catcode`\' = \active
\defŠ{\"a} \defš{\"o} \defŸ{\"u} \def€{\"A} \def…{\"O} \def†{\"U} \def§{\ss} \defŽ{\'{e}}
\def{\`{e}} \def'{\"{e}} \def™{\^{o}} \def{\^{e}} \def¿{\o} \def˜{\`{o}} \def'{\'{i}}

\begin{document}
\title{High-Contrast Interference in a Thermal Cloud of Atoms}
\author{D. E. Miller, J. R. Anglin, J. R. Abo-Shaeer, K. Xu, J. K. Chin, and W. Ketterle\footnote{Website: cua.mit.edu/ketterle\_group}}

\affiliation{Department of Physics, MIT-Harvard Center for
Ultracold Atoms, and Research Laboratory of Electronics, MIT,
Cambridge, MA 02139}
\date{\today}

\begin{abstract}
The coherence properties of a gas of bosonic atoms above the
 BEC transition temperature were studied.  Bragg diffraction was used
to create two spatially separated wave packets, which interfere
during expansion.  Given sufficient expansion time, high fringe
contrast could be observed in a cloud of arbitrary temperature.
Fringe visibility greater than $90\%$ was observed, which
decreased with increasing temperature, in agreement with a simple
model. When the sample was ``filtered" in momentum space using
long, velocity-selective Bragg pulses, the contrast was
significantly enhanced in contrast to predictions.

\end{abstract}


\maketitle

\begin{figure}[t]
\label{fig1}
\includegraphics[width=65mm]{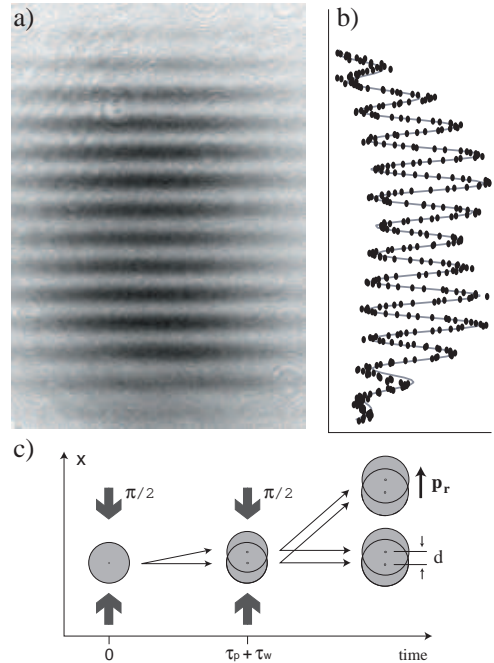}
\caption{Interference of spatially separated thermal clouds: (a)
An absorption image, taken after 48 ms time of flight, shows
fringes with a spatial period of  $\lambda_f = 340\ \mu m$. (b) A
cross section of the optical density (black circles) taken through
the center of the image was fit (grey line) to extract the fringe
contrast.  (c) Schematic depiction of the Bragg pulse sequence.
The first $\pi/2$ pulse at $t=0$ created a superposition of
stationary and moving clouds.  At $t=\tau_p+\tau_w$ the clouds had
separated by distance $d$, and a second pulse created a
superposition of cloud pairs; one pair moving with recoil momentum
$\mathbf{p}_r$, the other stationary. Each pair developed an
interference pattern.}
\end{figure}

Images of interfering atomic clouds are widely considered a
hallmark of Bose-condensed systems and a signature of long-range
correlations \cite{ketterle97}.  A thermal atomic cloud is often
regarded as an incoherent source, with a coherence length too
short to obtain high-contrast interference patterns when two
clouds are overlapped.  Here we show that ballistic expansion can
increase the coherence length such that ``BEC type" interference
can be observed in a thermal cloud.  Currently, there is
considerable interest in characterizing the coherence properties
of non-condensed systems including ultracold fermions
\cite{fermionReview}, fermion pairs
\cite{pairJin,pairZwierlein,pairGrimm} and ultra-cold molecules
\cite{herb03cs_mol,durrMol,xu03na_mol}.  In this paper we show
that an interferometric autocorrelation technique, previously only
applied to condensates \cite{phillips99,phillips00}, can be used
to study the coherence properties of samples at finite
temperature.


We studied the first-order spatial coherence of a trapped thermal
cloud of atoms. After release from the trap, the atom cloud
expanded ballistically, and Bragg diffraction was used to create
an identical copy of the initial cloud displaced by a distance
$d$. Therefore, our study was analogous to Young's double-slit
experiment \cite{carn91}.  We investigate the conditions under
which two such overlapping clouds produce a high-contrast
interference pattern. Our result is that for sufficiently long
expansion times, there will always be high contrast, but the
required time-of-flight becomes longer at higher temperatures.


The experiment used a magnetically trapped thermal cloud of $\sim
5 \times 10^7$ sodium atoms, prepared in a manner similar to our
previous work \cite{jamil01}. Atoms in the $\left| F=1, m_F=-1
\right>$ state were loaded from a MOT into a magnetic trap, where
they were further cooled by radio frequency (RF) evaporation.  The
RF evaporation was stopped before the critical temperature for
Bose-Einstein condensation (BEC) $T_c$ was reached, yielding a
thermal cloud at a controlled temperature. Shortly after being
released from the trap (2 ms), the cloud was exposed to two
successive Bragg pulses, separated by wait time $\tau_w$. The
effect of the Bragg beams \cite{pritchardBragg,lee95} was to
couple two momentum states $\left| \hbar\, \mathbf{k}_0 \right>$
and $\left| \hbar (\mathbf{k}_0 + \mathbf{k}_r) \right>$, via a
two photon transition, where $\mathbf{k}_r = (\mathbf{k}_1 -
\mathbf{k}_2)$ and $\mathbf{k}_1$, $\mathbf{k}_2$ are the wave
vectors of the two Bragg beams.  The coupling induced Rabi
oscillations and the pulse area was experimentally chosen to
correspond to $\pi/2$ (i.e., an atom originally in a well defined
momentum state was taken to an equal superposition of the two
states).  During the wait time $\tau_w$ the two states accrued
different phases before a second $\pi/2$ pulse mixed the states
again. Considering only two momentum states, this is equivalent to
Ramsey spectroscopy \cite{ramsey}.

\begin{figure}[t]
\label{emergence}
\includegraphics[width=85mm]{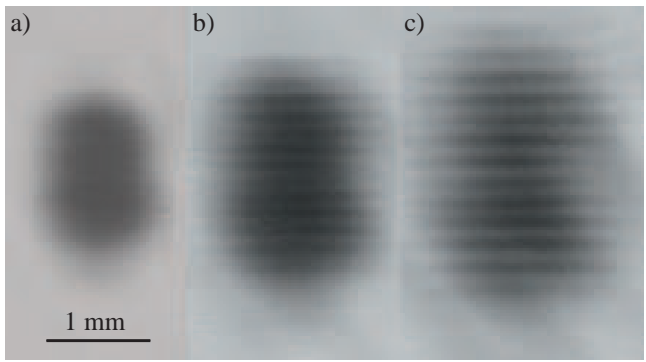}
\caption{Emerging contrast during ballistic expansion. The
coherence length $\ell_c$ grew larger than the initial separation
d=2 $\mu$m as the cloud expanded for (a) 14 ms, (b) 20 ms, (c) 25
ms.  (see Eq.\ \ref{coherence length})}
\end{figure}

One can regard a thermal cloud as an ensemble of atoms, each
having its initial amplitude spread over a range of momentum
states centered about zero, with r.m.s. width $h/\lambda_T$, where
$\lambda_T=h/\sqrt{mk_BT}$ is the thermal de Broglie wavelength
and $m$ the atomic mass.   The relative detuning of $\delta\nu =
45$ KHz between the Bragg beams and pulse duration $\tau_p=10\;\mu
s$ were chosen to couple all initial momentum components to those
centered at $\mathbf{k}_r$. During $\tau_w$ the relative phase
accumulated between coupled states is proportional to $k_0$, and
the momentum distribution shows a sinusoidal modulation (Ramsey
fringes)  after the Bragg sequence. In long time-of-flight (tof),
the spatial density simply mimics this momentum distribution.
Interference fringes were observed at a spacing $\lambda_f = h\,
t\,_{\rm{tof}}/m d$, where
\begin{equation} d = v_r\,\left(\tau_w +
4/\pi\,\tau_p\right) \end{equation} is the cloud separation
discussed in the equivalent picture of two overlapping clouds.
$v_r$ is the two-photon recoil velocity.  The factor of $4/\pi$
emerges from a simple Rabi oscillation model, and includes the
extra phase accumulated while the Bragg beams effect a $\pi/2$
pulse. The sum of single particle interference patterns results in
a density along the x-direction with reduced contrast C:
\begin{equation}
\label{density eqn} n(x) = f(x) \,\left[1+C\, \sin
\left(\frac{2\pi}{\lambda_f}x+\phi\right) \right].
\end{equation}
where $f(x)$ is an envelope function.

\begin{figure}[t]
\includegraphics[width=65mm]{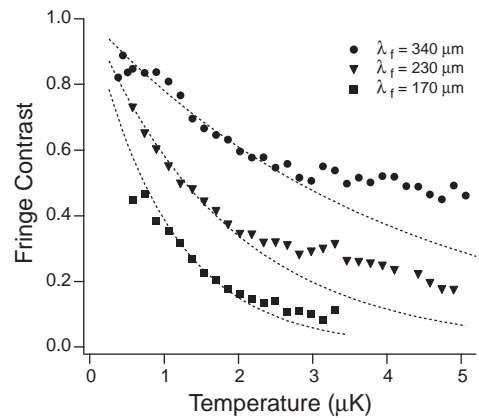}
\caption{Contrast vs.\ temperature for 10 $\mu$s Bragg pulses.
Different fringe spacings $\lambda_f$ were realized by varying the
wait time $\tau_w$ between the two Bragg pulses. The
time-of-flight was kept fixed at 48 ms.  At lower temperatures the
observed contrast agreed well with theory (dashed lines) given by
Eq.\ \ref{contrast}.  At higher temperature, however, the contrast
was higher than predicted, owing to the velocity selectivity of
the Bragg pulses. The BEC transition temperature was at $T_c=0.6\;
\mu K$.} \label{10uS graphs}
\end{figure}

To find the expected contrast we consider the $k$-space
distribution of noninteracting bosons in a harmonic trap assuming
the high temperature (Maxwell-Boltzmann) limit. This may be
written exactly as an ensemble of single-particle Gaussian states,
incoherently averaged over their positions.  Thus, it is
appropriate to regard the thermal cloud as a collection of wave
packets of gaussian width $\lambda_T$, distributed in space
according to Maxwell-Boltzmann statistics.

The incoherent sum over all particles results in contrast
\begin{equation} \label{contrast} C = \exp\left( -\frac{2 \pi^2 R_T^2}{\lambda_f^2}
\right) = \exp\left( -\frac{d^2}{2\,\ell_c^2} \right)
\end{equation}
\noindent where $R_T = \sqrt{k_BT/m\omega^2}$ is the thermal size
of a cloud in a harmonic trap of frequency $\omega$ and where we
have defined the coherence length as
\begin{equation}    \label{coherence length}
\ell_c = \frac{\lambda_T}{2 \pi} \,\frac{\mathrm{size\, in\,
tof}}{\mathrm{size\, in\, trap}}.
\end{equation}

Eq.\ 3 gives two different, but equivalent criteria for the loss
of contrast.  While the interference pattern of a single-particle
quantum state (as well as that of a pure condensate) should always
show perfect contrast, the incoherent sum over a thermal cloud
washes out the fringe visibility.  Contrast is lost when the
single particle interference pattern is smeared out over an
initial size $R_T$ that is larger than the fringe spacing
$\lambda_f$. However, because the fringe spacing grows as the
cloud expands, contrast will always emerge with enough time of
flight. Alternatively, Eq.\ 3 states that interference is lost
when the separation $d$ of the two sources exceeds the coherence
length $\ell_c$.  Here it is important to note, that the coherence
length increases with time-of-flight (Eq.\ 4).  The coherence
length is inversely proportional to the local momentum spread
which decreases in ballistic expansion as atoms with different
velocities separate from each other. For very long expansion
times, the coherence length becomes arbitrarily large resulting in
high-contrast interference (Fig.\ 2).  This can also be understood
by the conservation of local phase space density during ballistic
expansion, where the decrease in density is accompanied by a
decrease in momentum spread.

\begin{figure}[t]
\label{velocityselective}
\includegraphics[width=60mm]{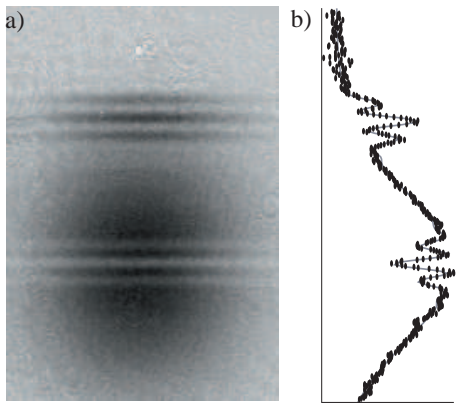}
\caption{Velocity selective Bragg diffraction:  (a) Absorption
image of atoms subject to 40 $\mu$s Bragg pulses. The fringe
spacing was $\lambda_f = 210\ \mu m$ after 48 ms time-of-flight.
These longer Bragg pulses addressed only a subset of the momentum
distribution. (b) Cross section and fit (black circles and grey
line respectively).}
\end{figure}


We repeated the experiment over a range of temperatures for
several values of the cloud separation $d$.  The temperature was
controlled by varying the final value of the RF evaporation.  Each
temperature was calibrated by measuring the size of a cloud in
expansion without pulsing on the Bragg beams.  The temperature
calibration was consistent with the observed onset of BEC at the
calculated temperature of $T_c=0.6\ \mu K$. The Bragg beams were
detuned 30 GHz from the atomic transition and heating was
demonstrated to be negligible by observing the effects of the
light with the two-photon detuning $\delta\nu$ set far from the
Bragg resonance. Absorption images of our samples were taken after
48 ms time-of-flight (Fig.\ 1a). A cross section of the atomic
density was fit to Eq.\ \ref{density eqn} to determine the
contrast. Deviation of the pulse area from $\pi/2$ reduces the
number of atoms in the out-coupled cloud, however it does not
reduce the contrast assuming the two pulses are equal.

Fig.\ 3a shows the measured fringe contrast of the out-coupled
cloud as a function of the cloud temperature for three different
fringe spacings $\lambda_f$.  The data is compared to the contrast
expected from Eq.\ \ref{contrast} (dashed line) with no free
parameters. At low temperature this equation provides an accurate
description of the observed fringe visibility.  At higher
temperatures, however, the observed contrast is consistently
greater than expected.  While Bose-Einstein statistics can
considerably enhance contrast at temperatures even above $T_c$
\cite{bloch00,glauber}, our experiment does not achieve the high
densities necessary to make this effect pronounced.  One possible
explanation for the discrepancy in Fig.\ 5 is that our model fails
when the Bragg diffraction becomes velocity selective, i.e., when
the Doppler width of the atoms exceeds the Fourier width of the
Bragg pulses.

\begin{figure}[t]
\label{velocityselectivedata}
\includegraphics[width=65mm]{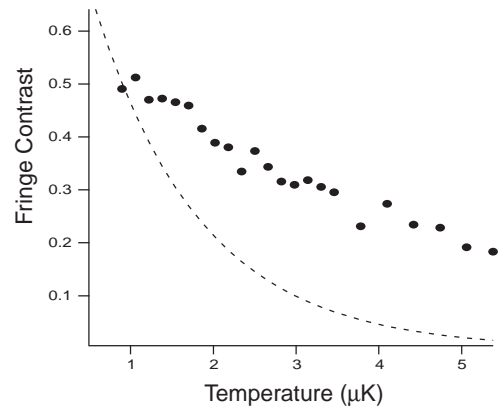}
\caption{Measured contrast for $\tau_p = 30~\mu s$ velocity
selective Bragg pulses.  The contrast was significantly higher
than that predicted by Eq.\ \ref{contrast} (dashed line). The
fringe spacing was $\lambda_f = 170\ \mu m$ after 48 ms expansion
time.}
\end{figure}

In order to further investigate this effect, we repeated the
experiment with longer, more velocity-selective pulses.  The
absorption images showed that these pulses addressed a narrow
range of atomic velocities (Fig.\ 4). In this case the fit routine
was modified to account for the (Gaussian) background of atoms
unaffected by the Bragg pulses. While the fraction of out-coupled
atoms decreased with increasing temperature, the interference was
still clearly visible. In Fig.\ 5 the measured contrast deviates
substantially from theory (dashed line). While velocity
selectivity clearly plays a role, the mechanism for this enhanced
contrast is not apparent. We have shown that it is only the
initial cloud size $R_T$ that determines the contrast for a given
$\lambda_f$ (Eq.\ \ref{contrast}).  Since the Bragg pulses are not
spatially selective, we do not expect their details to influence
the contrast.  Similarly, Eq.\ \ref{coherence length} illustrates
why velocity selection does not increase the coherence length in
time-of-flight: the narrowed momentum distribution implies a
larger effective de Broglie wavelength which is exactly cancelled
by the reduced expansion in time-of-flight.  Therefore, the
enhanced contrast in Fig.\ 5 cannot be described by the
single-particle free expansion of a thermal gas. We suspect that
particle interactions play a role.

In conclusion, we have shown how the coherence length of a trapped
gas is modified during ballistic expansion. This allows for the
observation of high-contrast interference in a sample at any
temperature, and is not limited to condensates.  For the technique
employed here, the only advantage of a BEC is its larger initial
coherence length, which is equal to the size of the condensate
\cite{ketterle97,phillips99,bloch00}.  However, it is the ability
of two independent condensates to interfere that sets this state
of matter apart \cite{Zoller96}.  Two independent thermal clouds
would not interfere. The self-interference technique characterized
here can be used to study the coherence properties of other novel
quantum degenerate systems. One example are molecular clouds
created by sweeping an external magnetic field across a Feshbach
resonance. While the rapid decay of such a sample precluded
thermalization, an interferometric method has already been used to
demonstrate coherence in this system \cite{moloptics}.

The authors thank D.\ E.\ Pritchard for critical reading of the
manuscript,  M.\ W.\ Zwierlein for contributions to the data
analysis and W.\ Setiawan for experimental assistance. This
research is supported by NSF, ONR, ARO and NASA.

\bibliographystyle{apsrev}
\bibliography{Refs}

\end{document}